\documentclass{article}
\usepackage{ijcai22}

\usepackage{footmisc}
\usepackage{times}
\usepackage{soul}
\usepackage{url}
\usepackage[hidelinks]{hyperref}
\usepackage[utf8]{inputenc}
\usepackage[small]{caption}
\usepackage{graphicx}
\usepackage{amsmath}
\usepackage{amsthm}
\usepackage{booktabs}
\usepackage{algorithm}
\usepackage{algorithmic}
\usepackage{makecell}
\usepackage{multirow}
\usepackage{pdfpages}

\newtheorem{example}{Example}

\usepackage{centernot}
\usepackage{array}
\usepackage{multirow}
\usepackage{amsfonts}
\usepackage{xcolor}
\usepackage[utf8]{inputenc}
\usepackage{comment}
\usepackage{todonotes}
\usepackage{amsfonts, amsmath, amssymb, bbm}
\usepackage{xcolor}
\usepackage{tikz}
\usepackage{hyperref}
\usepackage{tabularx}
\usepackage{subcaption} 
\usepackage{xpatch}
\usepackage{lscape}
\usepackage[misc,geometry]{ifsym}
\usepackage{lipsum}
\usepackage[absolute,overlay]{textpos}

\newcommand\blfootnote[1]{%
  \begingroup
  \renewcommand\thefootnote{}\footnote{#1}%
  \addtocounter{footnote}{-1}%
  \endgroup
}

\newcount\Comments  
\usepackage{color}
\definecolor{darkgreen}{rgb}{0,0.5,0}
\definecolor{purple}{rgb}{1,0,1}
\newcommand{\kibitz}[2]{\ifnum\Comments=1\textcolor{#1}{#2}\fi}
\newcommand{\xingyu}[1]  {\kibitz{darkgreen}   {[Xingyu: #1]}}
\newcommand{\wei}[1]  {\kibitz{purple}   {[Wei: #1]}}

\title{A Hierarchical HAZOP-Like Safety Analysis for Learning-Enabled Systems}

\author{
Yi Qi$^\dag$
\and
Philippa Ryan Conmy$^\ddag$
\and
Wei Huang$^\dag$
\and
Xingyu Zhao$^\dag$
\And
Xiaowei Huang$^\dag$
\affiliations
$^\dag$Department of Computer Science, University of Liverpool, Liverpool, L69 3BX, U.K.\\
\affiliations
$^\ddag$Adelard Part of NCC Group, London, N1 7UX, U.K.
\emails
\{yiqi,w.huang23,xingyu.zhao,xiaowei.huang\}@liverpool.ac.uk, pmrc@adelard.com \\
}

\begin{document}
\begin{textblock*}{20cm}(1cm,1cm)
\textcolor{red}{Preprint accepted by the AISafety'22 Workshop at IJCAI'22. To appear in a volume of CEUR Workshop Proceedings (\url{http://ceur-ws.org/})}.
\end{textblock*}
\maketitle
\begin{abstract}
\blfootnote{Copyright \textcopyright~2022 for this paper by its authors. Use permitted under Creative Commons License Attribution 4.0 International (CC BY 4.0).}Hazard and Operability Analysis (HAZOP) is a powerful safety analysis technique with a long history in industrial process control domain. With the increasing use of Machine Learning (ML) components in cyber physical systems---so called Learning-Enabled Systems (LESs), there is a recent trend of applying HAZOP-like analysis to LESs. While it shows a great potential to reserve the capability of doing sufficient and systematic safety analysis, there are new technical challenges raised by the novel characteristics of ML that require retrofit of the conventional HAZOP technique. In this regard, we present a new Hierarchical HAZOP-Like method for LESs (HILLS). To deal with the complexity of LESs, HILLS first does ``divide and conquer'' by stratifying the whole system into three levels, and then proceeds HAZOP on each level to identify \textit{(latent-)hazards}, \textit{causes}, security \textit{threats} and \textit{mitigation} (with new nodes and guide words). Finally, HILLS attempts at linking and propagating the causal relationship among those identified elements within and across the three levels via both qualitative and quantitative methods. We examine and illustrate the utility of HILLS by a case study on Autonomous Underwater Vehicles, with discussions on assumptions and extensions to real-world applications. HILLS, as a first HAZOP-like attempt on LESs that explicitly considers ML internal behaviours and its interactions with other components, not only uncovers the inherent difficulties of doing safety analysis for LESs, but also demonstrates a good potential to tackle them.

\end{abstract}

\section{Introduction}
\label{sec_intro}

After initially developed to support the chemical process industries (by Lawley \cite{lawley1974operability}), Hazard and Operability Analysis (HAZOP) has been successfully and widely applied in the past 50 years. It is generally acknowledged to be an effective yet simple method to systematically identify safety hazards. HAZOP is a prescriptive analysis procedure designed to study the system operability by analysing the effects of any deviation from its design intent \cite{CRAWLEY201510}. A HAZOP does semi-formal, systematic, and critical examination of the process and engineering intentions of the process design. The potential for hazards or operability problems are thus assessed, and malfunction of individual components and associated consequences for the whole system can be identified \cite{DUNJO201019}.


In recent years, increasingly sophisticated mathematical modelling processes from Machine Learning (ML) are being used to analyse complex data and then embedded into cyber physical systems---so called Learning-Enabled Systems (LESs). How to ensure the safety of LESs has become an enormous challenge \cite{lane_new_2016,zhao_safety_2020,asaadi_quantifying_2020}. As LESs are disruptively novel, they require new and advanced analysis for the complex requirements on their safe and reliable function \cite{BKCF2019}. Such analysis needs to be tailored to fully evaluate the new character of ML \cite{alves_considerations_2018,burton_mind_2020}, making conventional methods including HAZOP and HAZOP-like variants (e.g., CHAZOP \cite{andow1991guidance} and PES-HAZOP \cite{burns_modified_1993} that are respectively introduced for computer-based and programmable electronic systems) obsolete. Moreover, LESs exhibit unprecedented complexity, while past experience suggests that HAZOP should be continuously retrofitted to accommodate more complex systems \cite{Pasman_Rogers_2016}, considering quantitative analysis frameworks \cite{ozog1987hazard,cozzani2007hazmat} and human factors \cite{aspinall2006hazops}. To the best of our knowledge, there is no HAZOP-like safety analysis dedicated for LESs that takes into account ML characters while preserving the simplicity and effectiveness of HAZOP (comparing to other conventional safety analysis methods \cite{sun2022comparison}), which motivates this research.

In this paper, we introduce a new Hierarchical HAZOP-Like method for LESs (HILLS). HILLS first stratifies the complex LESs into three levels---System Level, ML-Lifecycle Level and Inner-ML Level, then applies HAZOP separately on each level to identify safety elements of interest, namely \textit{causes}, \textit{mitigation}, \textit{hazards} (or \textit{latent-hazards} for latent levels that cannot directly lead to mishaps) and security \textit{threats}. When applying HAZOP on the ML related levels, we revise HAZOP to cope with ML characteristics, e.g., by introducing new ways of defining nodes and new sets of guide words. We also identify causes of hazards from the ML \textit{development process} (modelled by the ML-Lifecycle level) to reflect its data-driven nature (e.g., how data is collected, processed, etc).
Furthermore, we attempt to address the challenge of how to link and propagate those identified safety elements within and across three levels, then propose both qualitative and quantitative (an initial Bayesian Belief Network (BN) solution) methods to model the casual relationships. To examine the effectiveness and demonstrate the use case of HILLS, we finally conduct a case study on Autonomous Underwater Vehicles (AUVs), with discussions on assumptions adopted and extensions to real-world applications.  


The key contributions of this work include:

\textit{a)} A first HAZOP-like safety analysis for LESs that explicitly considers ML characters (including security threats and the data-driven nature in the development process) and reduces the complexity by hierarchical design.


\textit{b)} New considerations of dividing nodes in the system representation and a set of new guide words that adapt the traditional HAZOP method for lower levels regarding ML models. 


\textit{c)} A first attempt at linking/propagating identified \textit{causes}, \textit{mitigation}, \textit{(latent-)hazards} \& \textit{threats} across ML levels.


\textit{d)} Key challenges identified as a set of research questions that are generic to safety analysis for LESs in future research.

\section{Preliminaries: HAZOP}


HAZOP is an inductive hazard assessment method that is conducted by an expert team. It systematically investigates each element in the system with the goal to find the potential situation that could cause the element to pose hazards or limit the system’s normal operations.

There are four basic steps to perform the HAZOP:

\begin{itemize}
    \item Define the project scope/aims, and form the expert team.
    \item Identify system elements and model the system as a system representation.
    \item Consider possible deviation of operational parameters.
    \item Identify hazards, causes and mitigation solutions.
\end{itemize}

Once the four steps are completed, team members may generate additional safety requirements if necessary to mitigate or prevent the identified issues, leading to improvement of the system. More details are given for each step of HAZOP as what follows.

\subsubsection{Form HAZOP team}

To perform HAZOP, a team of specialists is formed according to the project scope and aims. These experts have extensive experience, expert knowledge and understand the overall procedures of the system deeply, such as operations, maintenance and engineering design.

\subsubsection{Identify system elements}
The HAZOP team will formally represent the system under study by identifying the elements. Each element is called a \textbf{Node}, representing an operational function. Then, nodes and interactions between nodes (e.g., data/control flows) collectively form the system representation under analysis.

\subsubsection{Consider deviations of operational parameters}
HAZOP assumes that a problem can only arise when there are some \textbf{Deviations} from the intent design. HAZOP searches for deviations in the system representation. 
The deviation on a node is expressed as the combination of \textbf{Guide Words} and process \textbf{Attributes} .

Each guide word is a short word to create the imagination of a deviation of the design/process intent. The most commonly used guide words are: no, more, less, as well as, part of, other than, and so on. Guide words provide a systematic and consistent means of brainstorming potential deviations to normal operations.
Each guide word has a specific meaning, e.g., \textit{no} means the complete negation of the design intention, \textit{early} means something occurred earlier than intended time.

Attributes are closely related to nodes, and are usually the subject of the action being performed. 
The definition of attributes relies on 
expert knowledge.

\subsubsection{Identify hazards, causes and mitigation}
Where the result of a deviation would be a danger to workers or to the production process, a potential problem is found. \textbf{Hazard (H)} is a source of potential damage, harm or adverse health effects on something/someone, while mishaps are damages or harms on something/someone. 
\textbf{Cause (C)} is the reasons why the deviation could occur. It is possible that several causes are identified for one deviation. \textbf{Mitigation (M)} helps to reduce the occurrence frequency of the deviations or to mitigate their consequences. 
Hazards, causes, and mitigation are usually assigned with their respective IDs. 

\section{Problem Statement}
\label{sec_preliminary}


Given HAZOP was not originally designed for LESs, inevitably new problems arise when attempting to apply HAZOP on LESs. These problems are formalized as a set of research questions (RQs) proposed in this section. We first present the \textit{rationale} behind those RQs (i.e., justification of how we have come to the RQs) and then articulate what would be the \textit{expected solution} to each RQ. 


\paragraph{RQ1: How to reduce the complexity of LESs so that HAZOP can be effectively applied to?}
HAZOP is a semi-formalised analytical method, used to identify the hazard scenarios of a defined process, and it has been successfully used on \textit{relatively simple} systems. When facing a complex system, HAZOP often cannot play its role well. LESs exhibit unprecedented complexity, rendering directly applying HAZOP to LESs infeasible. Therefore, we need to reduce the complexity in the system representation. A simple yet effective solution is by ``divide and conquer'', e.g., stratifying a complex system into multiple levels. In this regard, we believe a promising solution to RQ1 is to propose a hierarchical system representation, so that HAZOP can be effectively applied.



\paragraph{RQ2: How to define nodes in each level, especially for novel levels regarding ML?}
We assume that HAZOP can effectively handle a single level system representation, as we expect to introduce a hierarchical structure in the RQ1 solution. The second step of HAZOP is to divide nodes at each level (presuming we already have a group of experts as the HAZOP team). Past experience shows that division of nodes can be based on the functionalities of components in the system \cite{unknown}, so we may continue using such traditional method for those non-ML related levels. However, when there are ML components in the system under analysis, it is difficult for the traditional division method of nodes to be directly applied. Therefore, RQ2 is raised to explore the novel definition of ``functionalities'' at ML-related levels. 


\paragraph{RQ3: Will there be any new guide words related to ML?}
Guide word is one of the key compositions of a deviation. The team of experts is responsible for identifying guide words that fit the scope of their analysis, while common guide words used were \textbf{No}, \textbf{Less/More}, \textbf{Slower/Faster}, \textbf{Early/Late}, etc. However, the existing set of guide words is unproven for use in ML applications, so this RQ aims at determining the effectiveness and new meanings of known guide words for ML related levels, and checking whether there might be missing guide words. Although we expect most of the known guide words can still be applicable,
they might miss some deviations given the new characteristics of ML. Thus, prospective new guide words may be introduced.

\paragraph{RQ4: How to establish the relationship between identified safety elements across levels?}
For simplicity, HAZOP is expected to be applied separately to each level of a hierarchical system representation. Therefore, to get the safety analysis of the whole complex system, it is necessary to study the relationship between identified safety elements---namely causes, mitigation, hazards (and latent-hazards)---across different levels. Then, based on the nature of the relationship (e.g., causal or not, quantitative or qualitative, probabilistic or deterministic), proper formalism should be used to establish and express such relationship of those hazard analysis results collected from each level.

\section{Running Example}
\label{sec_the_exa}

\begin{figure}[htb]
	\centering
	\includegraphics[width=\linewidth]{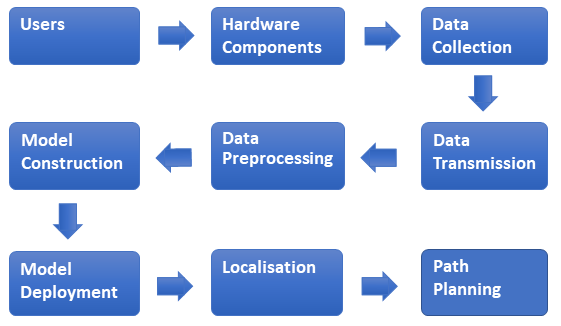}
	\caption{Workflow diagram of the running example}
	\label{fig_running_example}
\end{figure}

We present a running example from the SOLITUDE project\footnote{\url{https://github.com/Solitude-SAMR/UWV_RAM}\label{footnote1}}, which conducts safety analysis on an AUV that autonomously finds a dock and performs the docking task. The workflow of the scenario is given in Figure~\ref{fig_running_example}. 
The robot starts when received the user's command. Once started, it uses sensors (e.g., cameras) to receive data. Data is transmitted and preprocessed before feeding into the YOLO model for object detection and localisation. The localisation result is further utilised for path planning. In addition, the above normal workflow may suffer from external attacks on some stages, including data transmission, data preparation, and path planning. We remark that, the scenario in the project is more complex, including utilising deep reinforcement learning for motion planning, but for the space limit, this paper only focuses on the perception component. 

\section{Proposed Method}
\label{sec_the_met}

In this section, we present the HILLS method, and 
compare it with HAZOP.
HILLS is inheriting from HAZOP the
basic structure composition and definitions of elements, with extensions that are
suitable for LESs. The tables and figures presented in this section are partial for illustrative purpose only, cf. the complete HILLS analysis results based on the SOLITUDE project at the GitHub repository\footref{footnote1}.



\subsection{Hierarchical HAZOP}

\begin{figure}[ht]
	\centering
	\includegraphics[width=\linewidth]{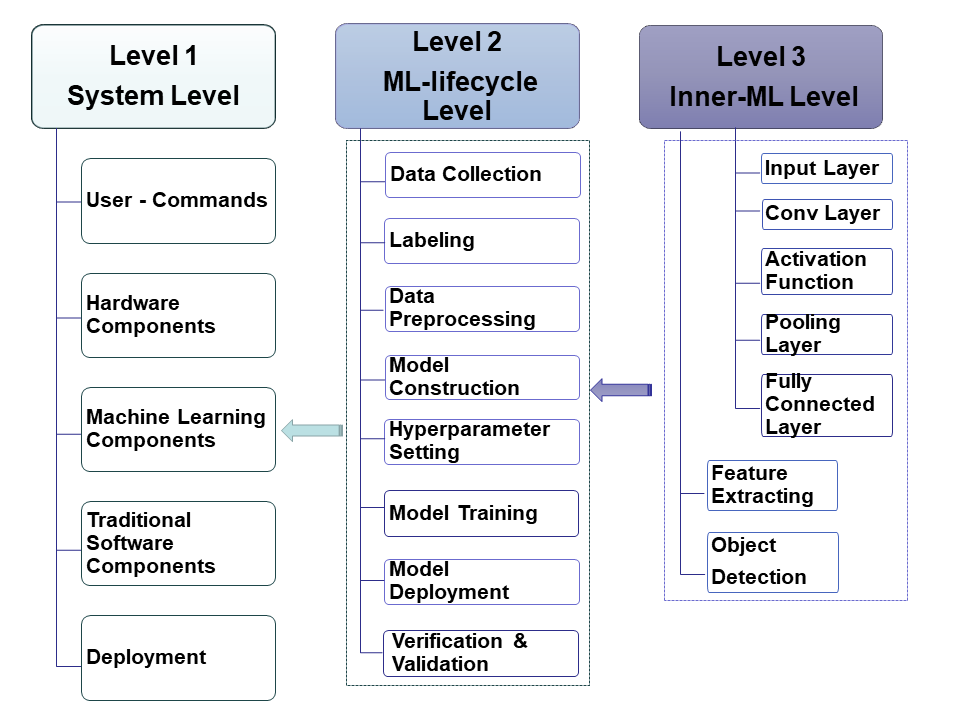}
	\caption{The 3 level hierarchical structure of HILLS}
	\label{fig_Method structure}
\end{figure}

As shown in Figure~\ref{fig_Method structure}, HILLS has a three-level structure, including \textit{system level}, \textit{ML-lifecycle level} and \textit{inner-ML level}. 
We analyse each level individually in this subsection, and
discuss their relations in Section~\ref{sec:relations}. Note, the HILLS structure discussed here is generic (for illustration purpose), and may be subject to adaptation when working with specific systems. 

\begin{table}[htb]
\caption{Nodes in each level in SOLITUDE example}\label{tab:nodesatlevels}
\centering
\setlength{\tabcolsep}{3mm}{
\begin{tabular}{@{}lll@{}}
\toprule
Level   & Node    & Description         \\ \midrule
System level & Node 1  & User                \\ 
System level & Node 2  & Hardware components     \\
System level & Node 3  & Data transmission   \\
ML-lifecycle level & Node 4  & Data collection \\
ML-lifecycle level & Node 5  & Labeling       \\
ML-lifecycle level & Node 6  & Data preprocessing \\
ML-lifecycle level & Node 7  & Hyperparameter setting \\
ML-lifecycle level & Node 8  & Model deployment          \\
Inner-ML level & Node 9  & Feature Extracting          \\
Inner-ML level & Node 10  & Object Detection           \\
ML-lifecycle level & Node 11  & Localisation       
       \\ \bottomrule
\end{tabular}%
}
\end{table}

\subsubsection{System level}

\begin{table*}[htb]
\caption{System level analysis (partial)}\label{systemlevel}
\centering
\resizebox{\textwidth}{!}{%
\begin{tabular}{@{}lllll@{}}
\toprule
Node            & Deviation       & Hazard               & Cause                          & Mitigation                                              \\ \midrule
Data transmission (Flow from camera to classifier)& No action      & Erratic trajectory       & No data from sensor (transient) & Acoustic guidance system                                            \\ 

Data transmission (Flow from camera to classifier)& No action      & Erratic trajectory        & No data from sensor (transient) & Situational awareness (route mapped and planned in advance)       \\
Data transmission (Flow from camera to classifier) & No action      & Erratic trajectory        & No data from sensor (transient) & Maximum safe distance maintained if uncertain                       \\ 
Data transmission (Flow from camera to classifier) & No action      & Insufficient energy/power & No data from sensor (permanent) & Camera health monitor (e.g. sanity check for blank images)          \\ \hline
Data transmission (Data flow)                      & Part of action & Erratic trajectory        & Corrupted sensor data           & Reliable camera (robust to environment etc.)                        \\ 
Data transmission (Data value)                    & Wrong value    & Loss of communication     & Hardware breakdown        & Hardware monitor               \\
Data transmission (Data value)                    & Wrong value    & Loss of communication     & Information conflict/lag        & Maximum safe distance maintained if uncertain                       \\ \bottomrule
\end{tabular}%
}
\end{table*}

HILLS at the system level largely follows HAZOP. 
Hardware, software, and ML components of an LES represent different functions, and they will be categorized as different \textbf{Nodes}.
Consider the running example in Figure~\ref{fig_running_example}, ``blue blocks'' represent the functional areas of the running example, which means that our nodes can be set according to these blocks. 
An example of setting nodes is provided in Table~\ref{tab:nodesatlevels}.
We note, the setting of nodes is specific to the system under investigation. E.g., the node ``Labeling'' was not included in Figure~\ref{fig_running_example}. 

Some {guide words} 
originated from, e.g., the chemical industry 
can still be used in LESs. {Attributes} related to the LES are 
used together with the guide words to express deviations. 
\begin{example}
{At system level, we discovered several hazards from the running example, some of them are summarised in Table~\ref{systemlevel}.
E.g., one of the hazards is ``erratic trajectory'', suggesting that the robot moves into an unsafe area. This hazard is associated with a deviation ``no action'' where ``no'' is the guide word and ``action'' is the attribute (when the AUV takes no actions in the water, the disturbance of current makes it difficult for the robot to maintain a stable trajectory). 
One of the causes of this hazard is ``no data from sensor'', which can be mitigated with, e.g., the deployment of an acoustic guidance system as a duplicated perception component based on another sensor. 
}
\end{example}


\begin{example}
Some hazards, such as ``erratic trajectory'', may appear in different nodes, which suggests that they may occur more often, and thus may have the higher priority to be mitigated after considering the severity of consequences as well.
\end{example}

\begin{example}
One hazard can be mitigated in different ways. For example, we identified several mitigation solutions for the ``erratic trajectory'', most of which focus on early prevention, such as ``maximum safe distance maintained if uncertain'' and ``camera health monitor''.
\end{example}

HILLS aims to exhaustively cover all potential hazards.
In the running example, 
the possible causes of crashes or failing to turn directions when facing obstacles may include ``no data from sensors (instantaneous or permanent)'', and ``misclassification'', corresponding to the errors in hardware and software components, respectively. However, the hazards, causes or mitigation may not be fully identifiable at this level. For example, there are other mitigation solutions for the cause ``misclassification'' that need to consider how the ML component is trained and constructed. However, the system level alone cannot naturally include  relevant nodes for this purpose. This motivates us to consider other levels (as discussed below).





\subsubsection{ML-Lifecycle level}


The key motivation for the ML-lifecycle level is to handle the complexity arising from the integration of ML components into an LES, considering mainly the human factors and security threats involved in the \textit{development process} of ML models. Thus, deviations from this level cannot be identified if analysis was only conducted at the system level.
On the other hand, the hazards at system level may be attributed to the hazards at ML-lifecycle level, e.g., the low prediction accuracy of ML component may be caused by the polluted data in the data collection or insufficient epochs of training.

For the running example, through the analysis at the ML-lifecycle level, we know that the low accuracy of the results may be caused by inaccurate labeling. 
We remark that, deviations identified at non system level are called \textbf{Latent-hazards (LH)}, as they pose \textit{indirect} hazards from latent levels with no hardware components being interacted and thus cannot directly lead to mishaps.






Table~\ref{redefine gw} presents a set of guide words that are required at this level. These guide words
are redefined from the existing guide words in HAZOP. Table~\ref{redefine gw} includes both their original meanings (in HAZOP) and new meanings (in HILLS). 
``part of'' represents a qualitative modification in the original meaning, and in HILLS it may mean the incompleteness of the structures, definitions, or settings. For ``Less'' and ``More'', considering that we are concerned about data flow and data value, their new meanings refer to the amount of data rather than, e.g., the water volume. 

\begin{table}[ht]
\caption{Redefined guide words in the ML-Lifecycle level}\label{redefine gw}
\centering
\begin{tabular}{@{}lll@{}}
\toprule
Guide word & Part of  \\                                    
Original meaning & There is a qualitative modification \\
New Meaning & Incomplete structure, definition or setting \\ \hline
Guide word & Less         \\
Original meaning & Too little water or additive volume added \\
New Meaning & A less amount of data \\ \hline
Guide word & More                            \\ 
Original meaning & Too much water or additive volume added \\
New Meaning & A large amount of data \\
\bottomrule
\end{tabular}%
\end{table}

Safety analysis at the ML-lifecycle level can exhibit new latent-hazards, as shown in Table~\ref{Ml level}. While ML models are subject to security issues, we believe malicious attacking behaviors
should also be considered as security \textbf{Threats (T)}. 

Human factors are considered because ML development is a human-centered process, which makes possible some human related errors such as labelling errors, part of operations were forgotten and the omission of data preparation. Aforementioned mistakes are direct human errors.
There are also adversarial attacks that can lead to significant drop in performance, which are classified as security threats. A few examples are shown in Table~\ref{Ml level}.


%
\begin{example}
On the node ``data collection'', there is a threat ``data poisoning'', which occurs because the input data is contaminated. A suggested mitigation is to deploy a detector based on data provenance. 
\end{example}


\begin{example}
For ML components, we identified mitigation, e.g., ``classifier reliability for critical objects \textgreater X'' \cite{zhao_assessing_2021}, to reduce misclassifications with safety impacts. 
\end{example}

\begin{example}
For the latent-hazards ``low prediction accuracy'', its  causes include ``users make mistakes on labelling'', ``data itself is missing'', and ``data itself is incomplete'', each of which has their suggested mitigation (cf. Table~\ref{Ml level}). 
\end{example}





\begin{table*}[htb]
\caption{ML-lifecycle level analysis (partial)}\label{Ml level}
\centering
\resizebox{\textwidth}{!}{%
\begin{tabular}{@{}lllll@{}}
\toprule
Node                    & Deviation         & Latent-hazard \& Threat                                           & Cause                                               & Mitigation                                                         \\ \midrule
Labeling (Manually label data) & Wrong label      & Low prediction accuracy                     & Users make mistake with labeling                   & Keep classifier accuracy/reliability for critical objects \textgreater X \\
Labeling (Manually label data) & Wrong label       & Low prediction accuracy                      & Users make mistake with labeling                     & Sanity check for ground truth and label attribute                   \\
Labeling (Manually label data) & Incapable label             & Low prediction accuracy                      & Data itself is incomplete                          & Keep classifier accuracy/reliability for critical objects \textgreater X \\
Labeling (Manually label data) & Incapable label             & Low prediction accuracy                      & Data itself is incomplete                            & Sanity check for ground truth and label attribute            \\ \hline
Data collection      & Attacked               & Data Poisoning                                     & Input data is contaminated                          & Detection based on data provenance                                  \\
Data preprocessing      & Part of data washing & Incorrect data ranges                    & Data washing incomplete                             & Consistency Check (e.g. Value range)                                \\
 \hline
Hyperparameter setting   & Wrong setting           & Inappropriate hyperparameter        & User make mistake with setting                           & Sanity check to hyperparameter                                                       \\
Hyperparameter setting   & Wrong setting           & Inappropriate hyperparameter       & Unsuitable hyperparameter for setting                           & Continuing monitor to hyperparameter                                                       \\ \hline
Model deployment       & Attacked               & Robustness Attacks                                    & Insert a calculated disturbance into the input data & Defensive Distillation                                              \\
Model deployment       & Attacked               & Backdoor                                    & Insert disturbance into the input data & XAI explain to input \\ \hline
Localisation             & No Localisation      & Lose estimation of position         & Hardware (sensors) breakdown                    & Situational awareness (route mapped and planned in advance)         \\
Localisation             & No Localisation      &  Lose estimation of position         & Hardware mismatch                     & Common time to synchronise data and results                         \\
Localisation             & Wrong Localisation   & Misposition         & Slip rate too large                     & Situational awareness (route mapped and planned in advance)         \\
Localisation             & Wrong Localisation   & Misposition         & Combination miss between hardware and ML                    & Common time to synchronise data and results
\\ \bottomrule
\end{tabular}%
}
\end{table*}

\begin{example}
There is a deviation ``attack'', whose threats are various attacks, e.g., evasion attack, backdoor attack, and data poisoning attack. Their respective cause is usually that a certain entity in the training or inference of an ML model (e.g., input instance, model structure, training, dataset) is perturbed, modified, or  contaminated. Their respective mitigation can be very specific (cf. Table~\ref{Ml level}), e.g., the backdoor detector in \cite{huang_embedding_2022} for tree ensemble classifiers.
\end{example}




\subsubsection{Inner-ML level}

ML components such as YOLO are composed of one or more ML models, each of which is formed of a set of functional layers. 
Even after a thorough analysis of all possible deviations (with mitigation solutions) in the ML development process modelled by our ML-lifecycle level, the ML components may not perform as expected, e.g., the convolutional layers fail to extract features accurately, and the fully connected layers fail to make reliable classifications. Thus, safety analysis on the internal structure of an ML component is required. 
%

%
%


At the inner-ML level, HILLS takes the method of extracting basic layers of an ML component to form a model for analysis. To cater for different complexity of the ML component, two extraction methods are proposed.
%
The first one deals with simple models with up to 5 layers. It follows the layer structure and considers each layer to represent a separate functionality. Consequently, each layer is defined as a node in the system representation.
%
%
The second one deals with more complex, larger models by abstracting a model into several functional \textit{blocks} and every block may contain a number of layers.
Our analysis in the running example follows the second method. 




\begin{table}[h]
\centering
\caption{New guide words of ML-Lifecycle and inner-ML levels}\label{new gw}
\begin{tabular}{@{}ll@{}}
\toprule
Guide words & \multicolumn{1}{c}{Meaning} \\ \midrule
Wrong & Wrong setting or data value \\
Invalid & \makecell*[l]{Invalid data value or data flow, possibly\\  conflicting with other components} \\
Incomplete & Incomplete data value \\ 
Perturbed & \makecell*[l]{Data was perturbed by external attackers}\\ 
Incapable & Part of data can not be labeled
 \\ \bottomrule
\end{tabular}
\end{table}

We identified several new guide words, as shown in Table~\ref{new gw}, which are highly relevant to the setup of the ML component and data flow. It is worth noting that the ``Perturbed'' is a special guide word that is needed when considering the existence of an external attacker. 
\begin{example} Deviations containing ``perturbed'' are usually proprietary attacks, e.g., we record ``perturbed dataset'' as ``attack'' and the threat as ``data poisoning'' (cf. Table~\ref{Ml level}).
\end{example}

\begin{table*}[]
\caption{Inner-ML level analysis (partial)}\label{Inner level}
\centering
\resizebox{\textwidth}{!}{%
\begin{tabular}{@{}lllll@{}}
\toprule
Node      & Deviation     &  Latent-hazard \& Threat                              & Cause                                              & Mitigation                              \\ \midrule
Feature extracting & Imprecise extracting     & Wrong outputs & Less layers                             & Using deeper layers  \\
Feature extracting & Wrong extracting     & Wrong outputs & Wrong hyperparameter setting                              & Using Explainable AI (XAI) to locate\\
Feature extracting & Wrong extracting & Wrong outputs & Unsuitable kernel size setting & Kernel size need to match dataset size            \\
Feature extracting & Wrong extracting & Dying ReLU problem & Learning rate setting too large &  Choosing suitable learning rate for ReLU (activation function) \\
Feature extracting & Wrong extracting & Losing information of figures & Unsuitable parameter setting in pooling layer & Evaluate whether need pooling layer \\ 
Feature extracting & Wrong extracting & Losing information of figures & Unsuitable parameter setting in pooling layer &Choose an appropriate pooling type \\
\bottomrule
\end{tabular}%
}
\end{table*}

As shown in Table~\ref{Inner level}, HILLS performs analysis inside an ML model, which in general is closely related to the internal structure of the model. 

\begin{example}
When the ML component has wrong output, we can get from the inner-ML level analysis that this may be related to the setting of the hyperparameter. Explainable AI (XAI) methods may help users to, e.g., locate which layer of neurons contribute the most to the wrong ML behaviours \cite{bach2015pixel} and detect backdoors \cite{zhao_baylime_2021}.
\end{example}
\begin{example}
At the inner-ML level, we focus on the ML model structure itself. E.g., unsuitable parameter setting in activation functions or pooling layers also make specific latent-hazards. It also leads to wrong outputs or losing part of information of figures (cf. Table \ref{Inner level}).
\end{example}

\subsubsection{Further Considerations on Use Cases of HILLS}

HAZOP is to provide a 
systematic, critical examination of the process (and engineering intent) of a new or existing facility, and should normally be done before the system is officially put into service \cite{10.1007/s10664-013-9277-5}. Nevertheless, we believe that HILLS can still be applied after the occurrence of an accident, in particular the recent technologies have enabled the recording of system executions through, e.g., direct observation, recorded video, or snapshot images. HILLS may use the recordings to identify related causes and hazards. 

Moreover, we note the following points when 
using HILLS.
%
First, when dealing with an LES, 
we focus on the workflow or the pipeline diagram of the entire system, to identify nodes according to the method we explained earlier. The analysis at the system level can help us identify the hazards sourced from the ML components, to enable the analysis at the lower levels. 
%
Second, guide words will be combined with the attributes 
of each node to form deviations. This will proceed sequentially following the level structure of HILLS, i.e., the deviations at the system level will be identified first, followed by the ML-lifecycle level, and 
the inner-ML level.
Third, before looking for (latent-)hazards, causes, and mitigation at each level, we are based on a reasonable assumption that mitigation solutions of higher levels
are easier than lower levels. 
That said, HILLS may not need to be conducted at the inner-ML level, and can stop when all hazards are found and mitigated at other levels.

\subsection{Relations Between Levels} \label{sec:relations}

Up to now, we have identified the nodes, attributes, guide words, (latent-)hazards, threats, causes, and mitigation solutions for individual levels in the HILLS framework. We also notice that the relations between these elements can be very complicated. This calls for a formal analysis of the relations.  
While formalising the relations between levels is a significant challenge, and there might not be one best way, we propose to study them both qualitatively and quantitatively.

\subsubsection{Qualitative Analysis} 
Qualitative analysis studies the connections between levels,
with the guide words as entry points.
The guide words and the deviations may have the following connections.


First of all, the same guide words at a level have strong associations, even if they are combined with different attributes. Second, if a guide word is the same between different levels, the one in the higher level may contribute as the main reason for the latent-hazard of the lower level. 

\begin{example}
We use ``no'' as an example. We can get a deviation ``no action'' at the system level, and have the deviation ``no localisation'' in the ML-lifecycle level. Given they share the same guide word, we should consider whether the ``no localisation'' has a causality relation with the ``no action''.
\end{example}

Moreover, it is assumed that there is an inclusive relationship between the guide words of the higher level and lower level, such as ``no'' and ``part of'', or there are similar meanings, such as ``invalid'' or ``incompatible''.

{The existence of a guide word with an inclusive relationship suggests that for the latent-hazard found in the lower level, its cause may belong to the higher level.
}

\begin{example}
If we choose ``No action'' at system level and ``Part of definition'' at the ML-lifecycle level (e.g., images without defined labels), then we may establish an inclusive relationship between ``No'' and ``Part of''.
\end{example}

\begin{example}
We use ``invalid data value'' and ``incompatible data value'' as examples, ``incompatible data value'' may lead to the low accuracy of output or no results, it has a similar meaning with ``invalid data value''.
\end{example}
Selecting guide words is arguably a quite subjective activity that experts may use different guide words with similar semantics to identify the same cause. To this end, the proposed way of establishing relationships across levels can only cope with the ideal case in which identical guide words are used. Thus, alternative methods are still needed for other cases, which forms our important future work.


\subsubsection{Quantitative Analysis}
A BN is a graphical model that presents probabilistic relationships between a set of variables by determining causal relationships between them \cite{LEE20095880}. It is also a powerful tool for knowledge representation and reasoning under uncertainty, visually presenting probabilistic relationships between a set of variables \cite{cheng2002learning}. Actually, BN has already been used to study the relation between latent features learned by a deep neural network \cite{berthier2021abstraction}. 
While using BN to express relationship of elements is not a new idea in traditional safety analysis \cite{article2021,denney_towards_2011,zhao_new_2012}. 
We take the relationship between several elements at the ML-lifecycle level and the inner-ML level as an example to explore the possibility of using BN to represent it. 
This is an idea of quantitatively expressing relationships, since the higher level contains some abstract concepts, it is difficult to represent in variables. Even if we assume that abstract concepts are represented using variables, it is hard to present Conditional Probability Tables (CPTs) as a prerequisite for BN to start.
All parameters used to quantify BN must be obtained based on system background and expert knowledge. 

\begin{figure}[htb]
	\centering
	\includegraphics[width=\linewidth]{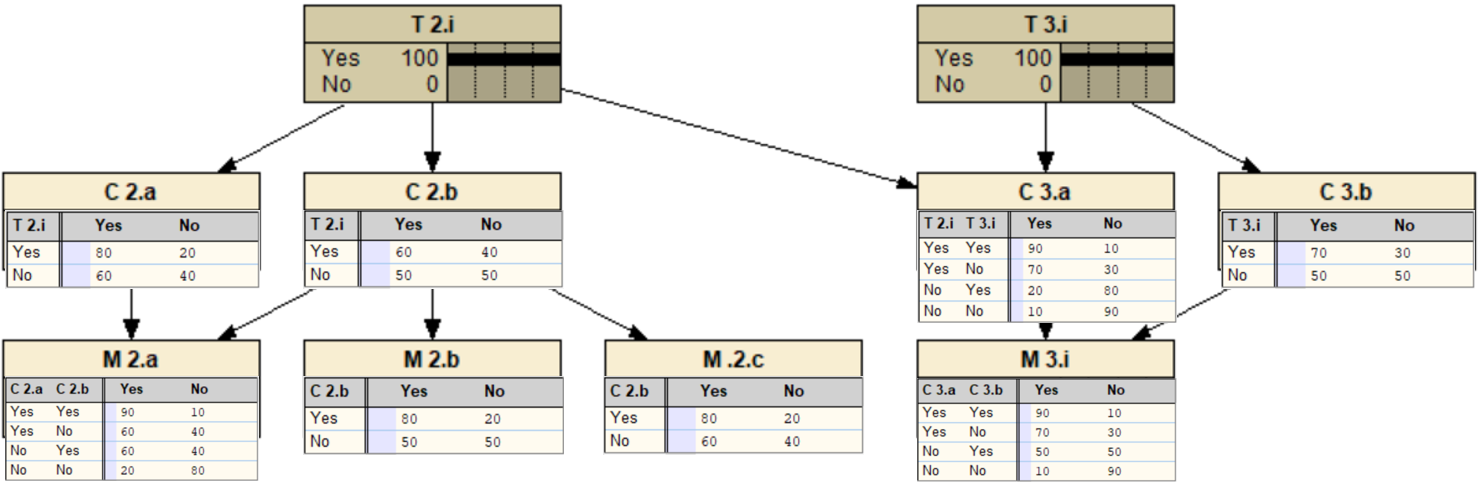}
	\caption{A BN model fragment (with illustrative probabilities)}
	\label{fig_BN}
\end{figure}

Figure~\ref{fig_BN} shows a fragment of the BN model for the running example, considering several security threats 
between the ML-lifecycle level and 
the inner-ML level. 
The nodes of a BN can represent threats ($Tl.i$), causes ($Cl.i$), or mitigation ($Ml.i$), where variable $l\in \{1,2,3\}$ ranges over the levels in HILLS and $i$ is the index of the threat/cause/mitigation at a level. E.g., $T2.i$ is the $i$-th threat at ML-lifecycle level. 

Besides, we need to assign CPT to each non-leaf node of the BN, and assign a prior probability to the leaf or set the observed evidence probability node.
It is noted that the expert knowledge is needed for both the construction of the basic structure and the assignment of CPTs. The probabilities used in Figure~\ref{fig_BN} are for illustrative purposes, while more enlightening examples can be found in \cite{berthier2021abstraction}. 

\begin{example}
For threat nodes with no incoming arrows, such as $T2.i$ and $T3.i$, we may set the probability of their occurrence 
to 100 percent. 
\end{example}

Once constructed, we can make probabilistic inference on the BN to ensure that the construction is correct w.r.t. expert knowledge. The following are two typical examples, by applying the d-separation algorithm \cite{koller2009probabilistic} (for determining dependencies of variables in a BN).

\begin{example}
There may be multiple children nodes at different levels for a parent node. In Figure~\ref{fig_BN}, the threat $T2.i$ has two causes, $C2.a$ and $C3.a$, at the ML-lifecycle level and inner-ML level, respectively. While the two causes may be mitigated separately as they belong to different levels, the effectiveness of their respective mitigation might affect the probabilistic inference based on each other's CPT (under the condition that the probability for $T2.1$ is not observable). 
\end{example}


\begin{example}
There may be multiple parent nodes for a child node. In Figure~\ref{fig_BN}, the mitigation $M2.a$, has two causes, $C2.a$ and $C2.b$, representing that one mitigation may support two causes. By observing the effectiveness of the mitigation (i.e., the CPT of $M2.a$), we will infer how one cause $C2.a$ may influence the other cause $C2.b$ and vice versa.  
\end{example}

We note, the construction of the BN structure and CPTs, as well as the above probabilistic inference, should be discussed and accepted by domain experts and all stakeholders. We believe BN is potentially a powerful tool for the purpose of modelling probabilistic causality relationship between elements of ML related levels, while how to apply BN in practice in the context of HILLS remains an open challenge.

\section{Related Work}
\label{sec_discussion}



\paragraph{HAZOP}
{
HAZOP is widely used in industrial domains, such as nuclear power~\cite{rimkevivcius2016hazop} and chemical industry~\cite{tian2015hazop}. In recent years, there has been efforts on integrating HAZOP with other methods ~\cite{marhavilas2020expanded,danko2019integration} to analyse common causes and system scenarios \cite{roche2019beyond}. 
A comprehensive review of those techniques may refer to recent survey papers, e.g. \cite{crawley2015hazop}. The application of HAZOP on computer-based systems first appears in \cite{chudleigh1992safety}. After that, the experience gained from application of HAZOP and related techniques to computer-based systems was summarised in \cite{KLETZ1997263}. There is a recent
trend of applying HAZOP-like analysis to LESs, e.g., in autonomous driving context \cite{kramer2020identification}.
}


\paragraph{Hierarchical structure} The concept of hierarchy is not new, but existing papers either focus on the hierarchical priority of the analysis order in the HAZOP analysis process \cite{othman2016prioritizing} or consider the direct application of the HAZOP to the hierarchical structure of traditional systems with no ML components \cite{nemeth2003hierarchical}.
A hierarchical structure is needed for its suitability to work with ML components (black-box in general, and inside the black-box, it is a layer-structure with each layer being a simple mathematical function). In HILLS, we innovatively consider the interaction between humans and ML components and the internal structure of the ML components. Moreover, inspired by \cite{WALLACE200553}, we investigate how to link and propagate identified safety elements at different levels.

\paragraph{STPA}
STAMP (Systems-Theoretic Accident Model and Processes) is also a very popular safety analysis method. STAMP uses three fundamental concepts from system theory: Emergence and hierarchy, communication and control, and process models \cite{leveson2011engineering}. STPA (System-Theoretic Process Analysis) uses such techniques, being based on the STAMP model. STPA pays more attention to the overall control loop and process analysis of the system, and focuses on unsafe
control actions and causal factors in a control structure. It is widely used in railway safety assurances \cite{YANG20191165}, cyber safety and security \cite{kaneko2018threat}, robotics \cite{ADRIAENSEN2021534} and driver-vehicle interactions \cite{9282848}. STPA is also used to explore a hierarchical structural safety analysis framework in \cite{CHAAL2020104939}. Comparing to STPA, HAZOP is relatively easier to conduct and clearer to communicate, supported by structural decomposition of the system functions \cite{sun2022comparison}. We start with retrofitting HAZOP for LESs, while STPA offers a new perspective to consider the feasibility of hierarchical safety analysis on LESs which is our planed future work.

\section{Conclusion}
\label{sec_conclusion}

We propose a hierarchical HAZOP-like method, HILLS, for the safety analysis of LESs. Being different from the traditional HAZOP, HILLS analyses LESs in a hierarchical way, disentangling the complexity by working with three separate levels first and then establishing their relations via both qualitative and quantitative methods, e.g., BNs. 
HILLS is applied to a practical example of AUVs, with the discovery of new guide words as well as new causes and mitigation related to ML. 
HILLS complements HAZOP when working with LESs, and is able to identify safety hazards and security threats related to ML components through its structural advantages. 





\section*{Acknowledgments}

This work is supported by U.K. DSTL through the project of Safety Argument for Learning-enabled Autonomous Underwater Vehicles and U.K. EPSRC through End-to-End Conceptual Guarding of Neural Architectures [EP/T026995/1]. \includegraphics[height=8pt]{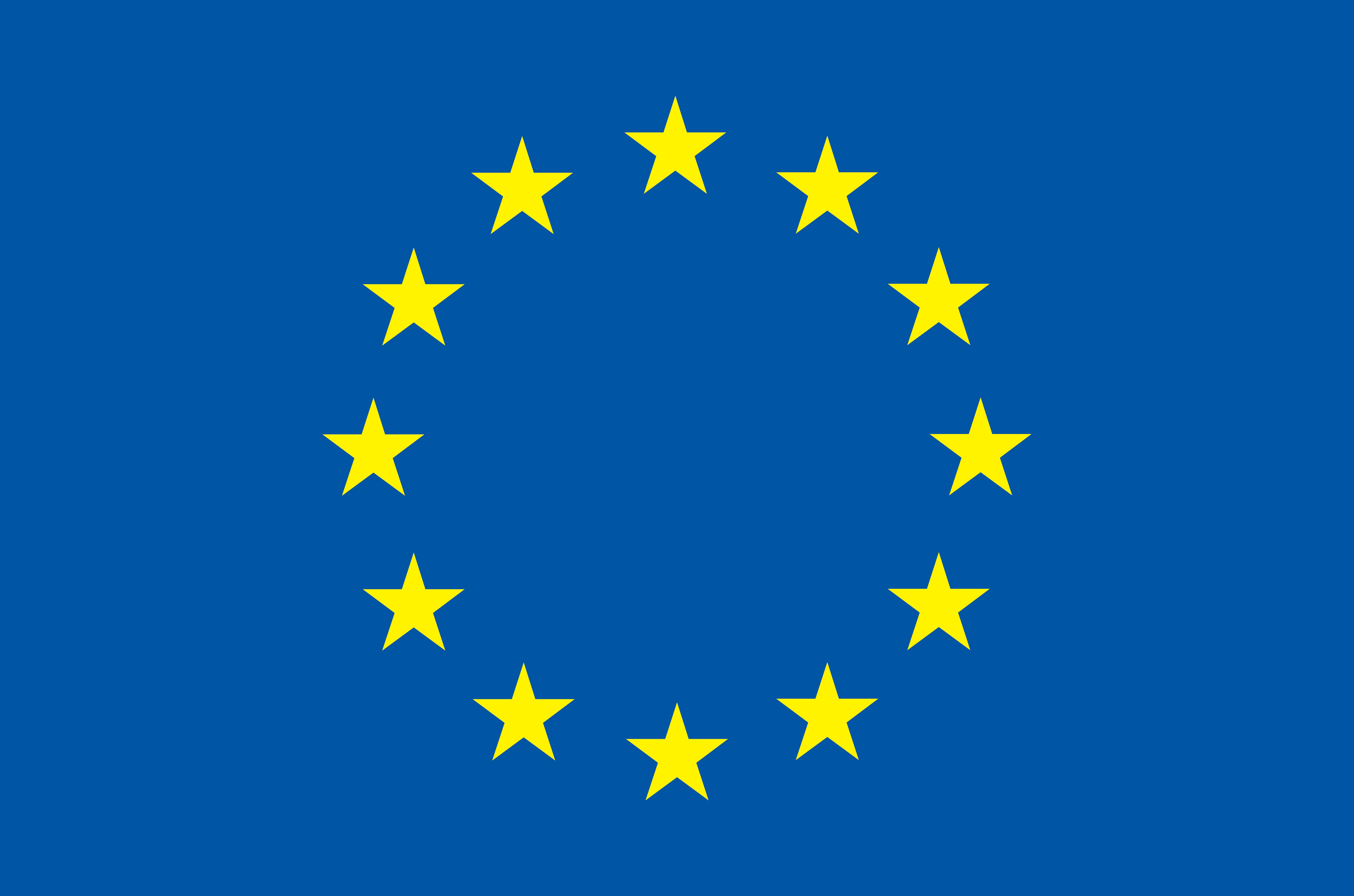} 
This project has received funding from the European Union’s Horizon 2020 research and innovation programme under grant agreement No 956123.
XZ’s contribution to the work is partially supported through Fellowships at the Assuring Autonomy International Programme. YQ’s contribution to the work
is supported through Chinese Scholarship Council (CSC).

\bibliographystyle{named}
\bibliography{references.bib}


\end{document}